# The Power Board of the KM3NeT Digital Optical Module: design, upgrade, and production

S. Aiello[a], A. Albert[b,bb], S. Alves Garre[c], Z. Aly[d], A. Ambrosone[f,e], F. Ameli[g], M. Andre[h], E. Androutsou[i], M. Anguita[j], L. Aphecetche[k], M. Ardid[l], S. Ardid[l], H. Atmani[m], J. Aublin[n], F. Badaracco[o], L. Bailly-Salins[p], Z. Bardačová[r,q], B. Baret[n], A. Bariego-Quintana[c], S. Basegmez du Pree[s], Y. Becherini[n], M. Bendahman[m,n], F. Benfenati[u,t], M. Benhassi[v,e], D.M. Benoit[w], E. Berbee[s], V. Bertin[d], V. van Beveren[s], S. Biagi[x], M. Boettcher[y], D. Bonanno[x], J. Boumaaza[m], M. Bouta[z], M. Bouwhuis[s], C. Bozza[aa,e], R.M. Bozza[f,e], H.Brânzaş[ab], F. Bretaudeau[k], R. Bruijn[ac,s], J. Brunner[d], R. Bruno[a], E. Buis[ad,s], R. Buompane[v,e], J. Busto[d], B. Caiffi[o], D. Calvo[c,*], S. Campion[g,ae], A. Capone[g,ae], F. Carenini[u,t], V. Carretero[c], T. Cartraud[n], P. Castaldi[af,t], V. Cecchini[c], S. Celli[g,ae], L. Cerisy[d], M. Chabab[ag], M. Chadolias[ah], C. Champion[n], A. Chen[ai], S. Cherubini[aj,x], T. Chiarusi[t], M. Circella[ak], R. Cocimano[x], J.A.B. Coelho[n], A. Coleiro[n], S. Colonges[n], R. Coniglione[x], P. Coyle[d], A. Creusot[n], G. Cuttone[x], R. Dallier[k], Y. Darras[ah], A. De Benedittis[e], B. De Martino[d], V. Decoene[k], R. Del Burgo[e], I. Del Rosso[u,t], U.M. Di Cerbo[e], L.S. Di Mauro[x], I. Di Palma[g,ae], A.F. Díaz[j], C. Diaz[j], D. Diego-Tortosa[x], C. Distefano[x], A. Domi[ah], C. Donzaud[n], D. Dornic[d], M. Dörr[al], E. Drakopoulou[i], D. Drouhin[b,bb], R. Dvornický[r], T. Eberl[ah], E. Eckerová[r,q], A. Eddymaoui[m], T. van Eeden[s], M. Eff[n], D. van Eijk[s], I. El Bojaddaini[z], S. El Hedri[n], A. Enzenhöfer[d], G. Ferrara[x], M. D. Filipović[am], F. Filippini[u,t], D. Franciotti[x], L.A. Fusco[aa,e], O. Gabella[an], J. Gabriel[an], S. Gagliardini[g], T. Gal[ah], J. García Méndez[l], A. Garcia Soto[c], C. Gatius Oliver[s], N. Geißelbrecht[ah], H. Ghaddari[z], L. Gialanella[e,v], B.K. Gibson[w], E. Giorgio[x], I. Goos[n], P. Goswami[n], D. Goupilliere[p], S.R. Gozzini[c], R. Gracia[ah], K. Graf[ah], C. Guidi[ao,o], B. Guillon[p], M. Gutiérrez[ap], H. van Haren[aq], A. Heijboer[s], A. Hekalo[al], L. Hennig[ah], J.J. Hernández-Rey[c], W. Idrissi Ibnsalih[e], G. Illuminati[u,t], P. Jansweijer[s], M. de Jong[ar,s], P. de Jong[ac,s], B.J. Jung[s], P. Kalaczyński[as,bc], O. Kalekin[ah], U.F. Katz[ah], A. Khatun[r], G. Kistauri[au,at], C. Kopper[ah], A. Kouchner[av,n], V. Kueviakoe[s], V. Kulikovskiy[o], R. Kvatadze[au], M. Labalme[p], R. Lahmann[ah], G. Larosa[x], C. Lastoria[d], A. Lazo[c], S. Le Stum[d], G. Lehaut[p], E. Leonora[a], N. Lessing[c], G. Levi[u,t], M. Lindsey Clark[n], P. Litrico[x], F. Longhitano[a], J. Majumdar[s], L. Malerba[o], F. Mamedov[q], J. Mańczak[c], A. Manfreda[e], M. Marconi[ao,o], A. Margiotta[u,t], A. Marinelli[e,f], C. Markou[i], L. Martin[k], J.A. Martínez-Mora[l], F. Marzaioli[v,e], M. Mastrodicasa[ae,g], S. Mastroianni[e], S. Miccichè[x], G. Miele[f,e], P. Migliozzi[e], E. Migneco[x], S. Minutoli[o], M.L. Mitsou[e], C.M. Mollo[e], L. Morales-Gallegos[v,e], M. Morga[ak], A. Moussa[z], I. Mozun Mateo[ax,aw], R. Muller[s], P. Musico[o], M.R. Musone[e,v], M. Musumeci[x], S. Navas[ap], A. Nayerhoda[ak], C.A. Nicolau[g], B. Nkosi[ai], B. Ó Fearraigh[ac,s], V. Oliviero[f,e], A. Orlando[x], E. Oukacha[n], D. Paesani[x], J. Palacios González[c], G. Papalashvili[ak,at], V. Parisi[ao,o], E.J. Pastor Gomez[c], A. M. Păun[ab], G.E. Păvălaş[ab], G. Pellegrini[t], S. Peña Martínez[n], M. Perrin-Terrin[d], J. Perronnel[p], V. Pestel[ax], R. Pestes[n], P. Piattelli[x], C. Poirè[aa,e], V. Popa[ab], T. Pradier[b], J. Prado[c], S. Pulvirenti[x], G. Quéméner[p], C.A. Quiroz-Rangel[l], U. Rahaman[c], N. Randazzo[a], R. Randriatoamanana[k], S. Razzaque[ay], I.C. Rea[e], D. Real[*,c], G. Riccobene[x], J. Robinson[y], A. Romanov[ao,o], A. Šaina[c], F. Salesa Greus[c], D.F.E. Samtleben[ar,s], A. Sánchez Losa[c,ak], S. Sanfilippo[x], M. Sanguineti[ao,o], C. Santonastaso[v,e], D. Santonocito[x], P. Sapienza[x], J. Schmelling[s], J. Schnabel[ah], J. Schumann[ah], H. M. Schutte[y], J. Seneca[s], N. Sennan[z], B. Setter[ah], I. Sgura[ak], R. Shanidze[at], A. Sharma[n], Y. Shitov[q], F. Šimkovic[r], A. Simonelli[e], A. Sinopoulou[a], M.V. Smirnov[ah], B. Spisso[e], M. Spurio[u,t], D. Stavropoulos[i], I. Štekl[q], M. Taiuti[ao,o], Y. Tayalati[m], H. Thiersen[y], I. Tosta e Melo[a,aj], E. Tragia[i], B. Trocmé[n], V. Tsourapis[i], E. Tzamariudaki[i], A. Vacheret[p], A. Valer Melchor[s], V. Valsecchi[x], V. Van Elewyck[av,n], G. Vannoye[d], G. Vasileiadis[az], F. Vazquez de Sola[s], C. Verilhac[n], A. Veutro[g,ae], S. Viola[x], D. Vivolo[v,e], J. Wilms[ba], E. de Wolf[ac,s], H. Yepes-Ramirez[l], G. Zarpapis[i], S. Zavatarelli[o], A. Zegarelli[g,ae], D. Zito[x], J.D. Zornoza[c], J. Zúñiga[c], N. Zywucka[y]

[a] *INFN, Sezione di Catania, Via Santa Sofia 64, Catania, 95123 Italy*
[b] *Université de Strasbourg, CNRS, IPHC UMR 7178, F-67000 Strasbourg, France*
[c] *IFIC - Instituto de Física Corpuscular (CSIC - Universitat de València), c/Catedrático José Beltrán, 2, 46980 Paterna, Valencia, Spain*
[d] *Aix Marseille Univ, CNRS/IN2P3, CPPM, Marseille, France*

[e] INFN, Sezione di Napoli, Complesso Universitario di Monte S. Angelo, Via Cintia ed. G, Napoli, 80126 Italy
[f] Università di Napoli "Federico II", Dip. Scienze Fisiche "E. Pancini", Complesso Universitario di Monte S. Angelo, Via Cintia ed. G, Napoli, 80126 Italy
[g] INFN, Sezione di Roma, Piazzale Aldo Moro 2, Roma, 00185 Italy
[h] Universitat Politècnica de Catalunya, Laboratori d'Aplicacions Bioacústiques, Centre Tecnològic de Vilanova i la Geltrú, Avda. Rambla Exposició, s/n, Vilanova i la Geltrú, 08800 Spain
[i] NCSR Demokritos, Institute of Nuclear and Particle Physics, Ag. Paraskevi Attikis, Athens, 15310 Greece
[j] University of Granada, Dept. of Computer Architecture and Technology/CITIC, 18071 Granada, Spain
[k] Subatech, IMT Atlantique, IN2P3-CNRS, Nantes Université, 4 rue Alfred Kastler - La Chantrerie, Nantes, BP 20722 44307 France
[l] Universitat Politècnica de València, Instituto de Investigación para la Gestión Integrada de las Zonas Costeras, C/ Paranimf, 1, Gandia, 46730 Spain
[m] University Mohammed V in Rabat, Faculty of Sciences, 4 av. Ibn Battouta, B.P. 1014, R.P. 10000 Rabat, Morocco
[n] Université Paris Cité, CNRS, Astroparticule et Cosmologie, F-75013 Paris, France
[o] INFN, Sezione di Genova, Via Dodecaneso 33, Genova, 16146 Italy
[p] LPC CAEN, Normandie Univ, ENSICAEN, UNICAEN, CNRS/IN2P3, 6 boulevard Maréchal Juin, Caen, 14050 France
[q] Czech Technical University in Prague, Institute of Experimental and Applied Physics, Husova 240/5, Prague, 110 00 Czech Republic
[r] Comenius University in Bratislava, Department of Nuclear Physics and Biophysics, Mlynska dolina F1, Bratislava, 842 48 Slovak Republic
[s] Nikhef, National Institute for Subatomic Physics, PO Box 41882, Amsterdam, 1009 DB Netherlands
[t] INFN, Sezione di Bologna, v.le C. Berti-Pichat, 6/2, Bologna, 40127 Italy
[u] Università di Bologna, Dipartimento di Fisica e Astronomia, v.le C. Berti-Pichat, 6/2, Bologna, 40127 Italy
[v] Università degli Studi della Campania "Luigi Vanvitelli", Dipartimento di Matematica e Fisica, viale Lincoln 5, Caserta, 81100 Italy
[w] E. A. Milne Centre for Astrophysics, University of Hull, Hull, HU6 7RX, United Kingdom
[x] INFN, Laboratori Nazionali del Sud, Via S. Sofia 62, Catania, 95123 Italy
[y] North-West University, Centre for Space Research, Private Bag X6001, Potchefstroom, 2520 South Africa
[z] University Mohammed I, Faculty of Sciences, BV Mohammed VI, B.P. 717, R.P. 60000 Oujda, Morocco
[aa] Università di Salerno e INFN Gruppo Collegato di Salerno, Dipartimento di Fisica, Via Giovanni Paolo II 132, Fisciano, 84084 Italy
[ab] ISS, Atomistilor 409, Măgurele, RO-077125 Romania
[ac] University of Amsterdam, Institute of Physics/IHEF, PO Box 94216, Amsterdam, 1090 GE Netherlands
[ad] TNO, Technical Sciences, PO Box 155, Delft, 2600 AD Netherlands
[ae] Università La Sapienza, Dipartimento di Fisica, Piazzale Aldo Moro 2, Roma, 00185 Italy
[af] Università di Bologna, Dipartimento di Ingegneria dell'Energia Elettrica e dell'Informazione "Guglielmo Marconi", Via dell'Università 50, Cesena, 47521 Italia
[ag] Cadi Ayyad University, Physics Department, Faculty of Science Semlalia, Av. My Abdellah, P.O.B. 2390, Marrakech, 40000 Morocco
[ah] Friedrich-Alexander-Universität Erlangen-Nürnberg (FAU), Erlangen Centre for Astroparticle Physics, Nikolaus-Fiebiger-Straße 2, 91058 Erlangen, Germany
[ai] University of the Witwatersrand, School of Physics, Private Bag 3, Johannesburg, Wits 2050 South Africa
[aj] Università di Catania, Dipartimento di Fisica e Astronomia "Ettore Majorana", Via Santa Sofia 64, Catania, 95123 Italy
[ak] INFN, Sezione di Bari, via Orabona, 4, Bari, 70125 Italy
[al] University Würzburg, Emil-Fischer-Straße 31, Würzburg, 97074 Germany
[am] Western Sydney University, School of Computing, Engineering and Mathematics, Locked Bag 1797, Penrith, NSW 2751 Australia
[an] IN2P3, LPC, Campus des Cézeaux 24, avenue des Landais BP 80026, Aubière Cedex, 63171 France
[ao] Università di Genova, Via Dodecaneso 33, Genova, 16146 Italy
[ap] University of Granada, Dpto. de Física Teórica y del Cosmos & C.A.F.P.E., 18071 Granada, Spain
[aq] NIOZ (Royal Netherlands Institute for Sea Research), PO Box 59, Den Burg, Texel, 1790 AB, the Netherlands
[ar] Leiden University, Leiden Institute of Physics, PO Box 9504, Leiden, 2300 RA Netherlands
[as] National Centre for Nuclear Research, 02-093 Warsaw, Poland
[at] Tbilisi State University, Department of Physics, 3, Chavchavadze Ave., Tbilisi, 0179 Georgia
[au] The University of Georgia, Institute of Physics, Kostava str. 77, Tbilisi, 0171 Georgia
[av] Institut Universitaire de France, 1 rue Descartes, Paris, 75005 France
[aw] IN2P3, 3, Rue Michel-Ange, Paris 16, 75794 France
[ax] LPC, Campus des Cézeaux 24, avenue des Landais BP 80026, Aubière Cedex, 63171 France
[ay] University of Johannesburg, Department Physics, PO Box 524, Auckland Park, 2006 South Africa
[az] Laboratoire Univers et Particules de Montpellier, Place Eugène Bataillon - CC 72, Montpellier Cédex 05, 34095 France
[ba] Friedrich-Alexander-Universität Erlangen-Nürnberg (FAU), Remeis Sternwarte, Sternwartstraße 7, 96049 Bamberg, Germany
[bb] Université de Haute Alsace, rue des Frères Lumière, 68093 Mulhouse Cedex, France
[bc] AstroCeNT, Nicolaus Copernicus Astronomical Center, Polish Academy of Sciences, Rektorska 4, Warsaw, 00-614 Poland

## Abstract

The KM3NeT Collaboration is building an underwater neutrino observatory at the bottom of the Mediterranean Sea consisting of two neutrino telescopes, both composed of a three-dimensional array of light detectors, known as digital optical modules. Each digital optical module contains a set of 31 three-inch photomultiplier tubes distributed over the surface of a 0.44 m diameter pressure-resistant glass sphere. The module includes also calibration instruments and electronics for power, readout and data acquisition. The power board was developed to supply power to all the elements of the digital optical module. The design of the power board began in 2013, and several prototypes were produced and tested. After an exhaustive validation process in various laboratories within the KM3NeT Collaboration, a mass production batch began, resulting in the construction of over 1,200 power boards so far. These boards were integrated in the digital optical modules that have already been produced and deployed, 828 until October 2023. In 2017, an upgrade of the power board, to increase reliability and efficiency, was initiated. After the validation of a pre-production series, a production batch of 800 upgraded boards is currently underway. This paper describes the design, architecture, upgrade, validation, and production of the power board, including the reliability studies and tests conducted to ensure the safe operation at the bottom of the Mediterranean Sea throughout the observatory's lifespan.

*Keywords:* power supply; acquisition electronics; neutrino telescopes.

## 1. Introduction

The KM3NeT Collaboration is building two underwater neutrino telescopes in the Mediterranean Sea for the detection of astrophysical neutrinos and the study of the fundamental neutrino properties measuring the oscillation patterns of atmospheric neutrinos [1]. These detectors, known as Astroparticle Research with Cosmics in the Abyss (ARCA) and Oscillation Research with Cosmics in the Abyss (ORCA), are located off the southern coast of Sicily, Italy, and near the coast of Toulon, France, respectively, at depths of approximately 3,500 m and 2,450 m. The telescopes are built in the form of 3D lattices of light detectors called Digital Optical Modules (DOMs) [2, 3], each containing 31 three-inch photomultiplier tubes (PMTs) [4], instrumentation for calibration and positioning and all associated electronics boards. The DOMs are used to reconstruct the trajectory and energy of the primary neutrino by measuring the arrival times and positions of the Cherenkov photons induced by the relativistic charged particles produced in the interaction of neutrinos with matter inside and nearby the telescopes. The designed instrumented volume is around 1 $km^3$ for ARCA and $7\times10^6$ $m^3$ for ORCA. The DOMs are distributed along lines called Detection Units (DUs), each containing 18 DOMs. The DUs are anchored on the seafloor and kept vertical by the buoyancy of the DOMs and buoys at the top. The horizontal spacing between DUs is approximately 90 m in ARCA and 20 m in ORCA, whereas the vertical spacing is around 36 m in ARCA and 9 m in ORCA. The different spatial configurations of ARCA and ORCA correspond to the different scientific scopes and neutrino energy ranges, ARCA being optimized for the detection of cosmic neutrinos (TeV − PeV) and ORCA for atmospheric neutrinos (1 GeV − 1 TeV). ARCA will consist of 230 DUs distributed in two separated blocks while ORCA will be formed by only one block of 115 DUs. As of October 2023, the number of DUs deployed was 28 in ARCA and 18 in ORCA. See Figure 1 for a sketch view of the KM3NeT detector.

The main data acquisition electronic board of the DOM is the Central Logic Board (CLB) [5, 6, 7], which mainly performs the readout of the 31 PMT channels. The Power Board (PB) is mounted on the bottom of the CLB and provides power to all the DOM elements. This includes generating the high voltage for the 31 PMTs, for all the instrumentation devices located inside the DOM: compass, accelerometer, gyroscope,

*corresponding author

*Email address:* km3net-pc@km3net.de; real@ific.uv.es (D. Real); dacaldia@ific.uv.es (D. Calvo)

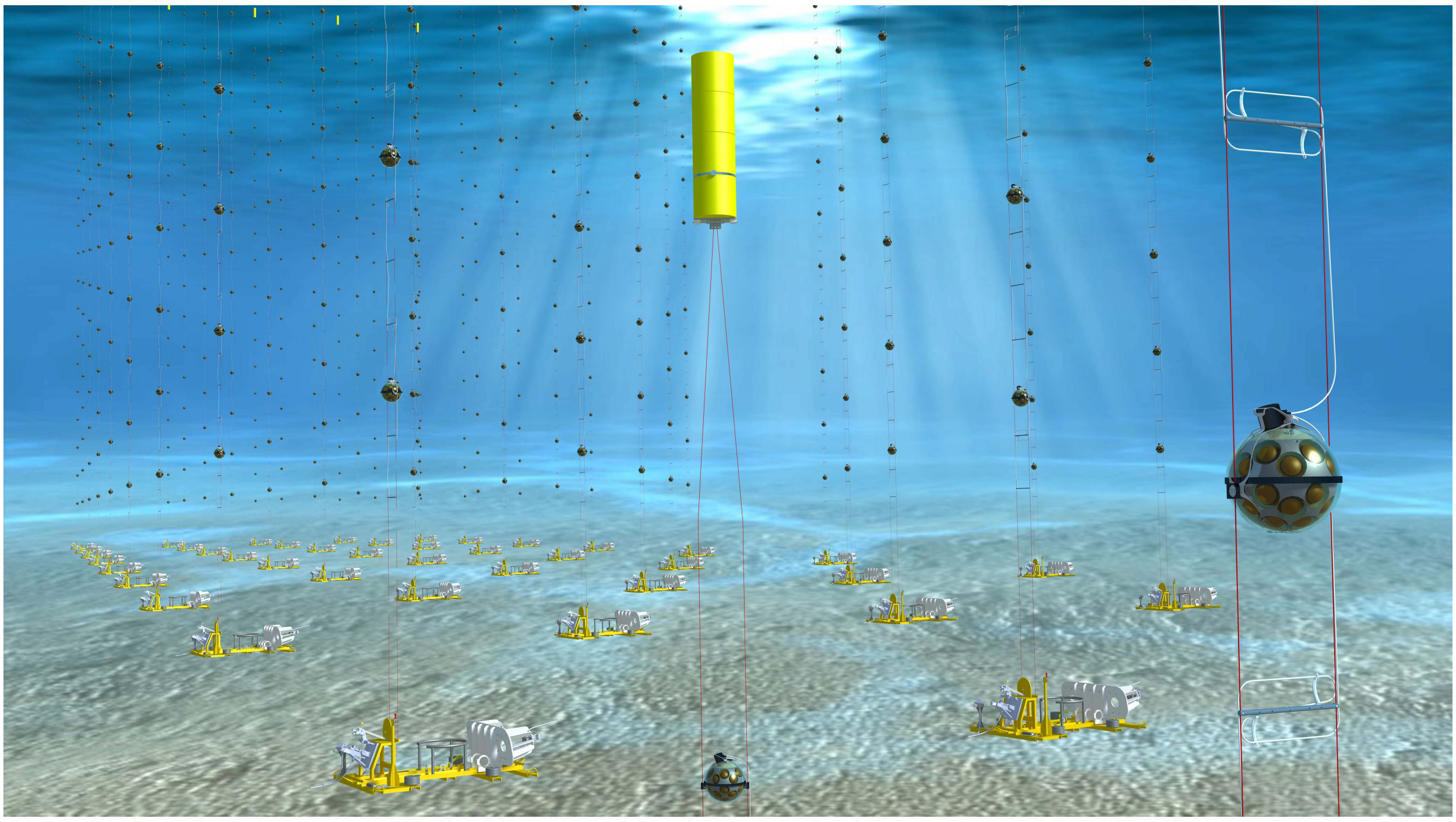

Figure 1: Artistic view of the KM3NeT detector. The illustration is not to scale: sunlight does not reach the depths at which the KM3NeT detector is deployed. The total instrumented volume of the KM3NeT detectors, once completed, will be around one km$^3$ for ARCA and $7{\times}10^6$ m$^3$ for ORCA.

pressure, humidity, and temperature sensors, and the power for the time and positioning calibration devices: the Nanobeacon flasher [8] and the acoustic piezo sensor.

The PB has to generate the different voltages for the Field Programmable Gate Array (FPGA) of the CLB in the proper startup sequence. The Xilinx FPGA needs a monotonic startup sequence for all the voltages to avoid increasing inrush current. The PB also includes the readout interface to monitor all the voltages and currents generated. To isolate the power circuit, reduce noise, and enhance reliability in the CLB, it was decided to build a separate power board in the KM3NeT acquisition electronics. This approach has also been adopted by other physics experiments such as ALICE [9] or ATLAS TileCAL [10].

The PB is located in the shielded part of the aluminum cooling frame in the DOM, to protect the sensitive electronics inside the DOM from interference caused by the high-frequency noise produced by the DC/DC converters of the PB. This location also provides better cooling for the PB. The frame at that point is shaped like a solid spherical cap that makes full contact with the inner surface of the DOM glass sphere. This helps maximize the heat flow to the surrounding seawater, where the ambient temperature is around 13 °C, acting as a heat sink for the thermal losses of the converters of the PB. Figure 2 shows the different elements inside the DOM as well as the location of the PB.

The initial design of the power board was completed in 2013. A pre-serie of prototypes were built and submitted to numerous and rigorous tests, such as HALT tests [11], efficiency tests, etc. After an extensive validation process across various laboratories within the KM3NeT Collaboration, more than 1,200 power boards were manufactured. These were integrated into the DOMs which have already been deployed in the sea, numbering 828 as of October 2023. An upgrade of the power board began in 2017 with the aim of improving efficiency and reliability. Following validation of the pre-production series, a production batch of 800 improved boards is currently being worked on.

This paper is organized as follows. The design and architecture of the KM3NeT power board are detailed in Section 2, while the upgrade of the PB, primarily through a better selection of converters to improve efficiency and reduce consumption, is presented in Section 3. Section 4 describes the analysis of reliability. The production, control and the functional tests applied are discussed in Sections 5 and 6, while in Section 7 the results on production and reliability are presented. Finally, conclusions are drawn in Section 8.

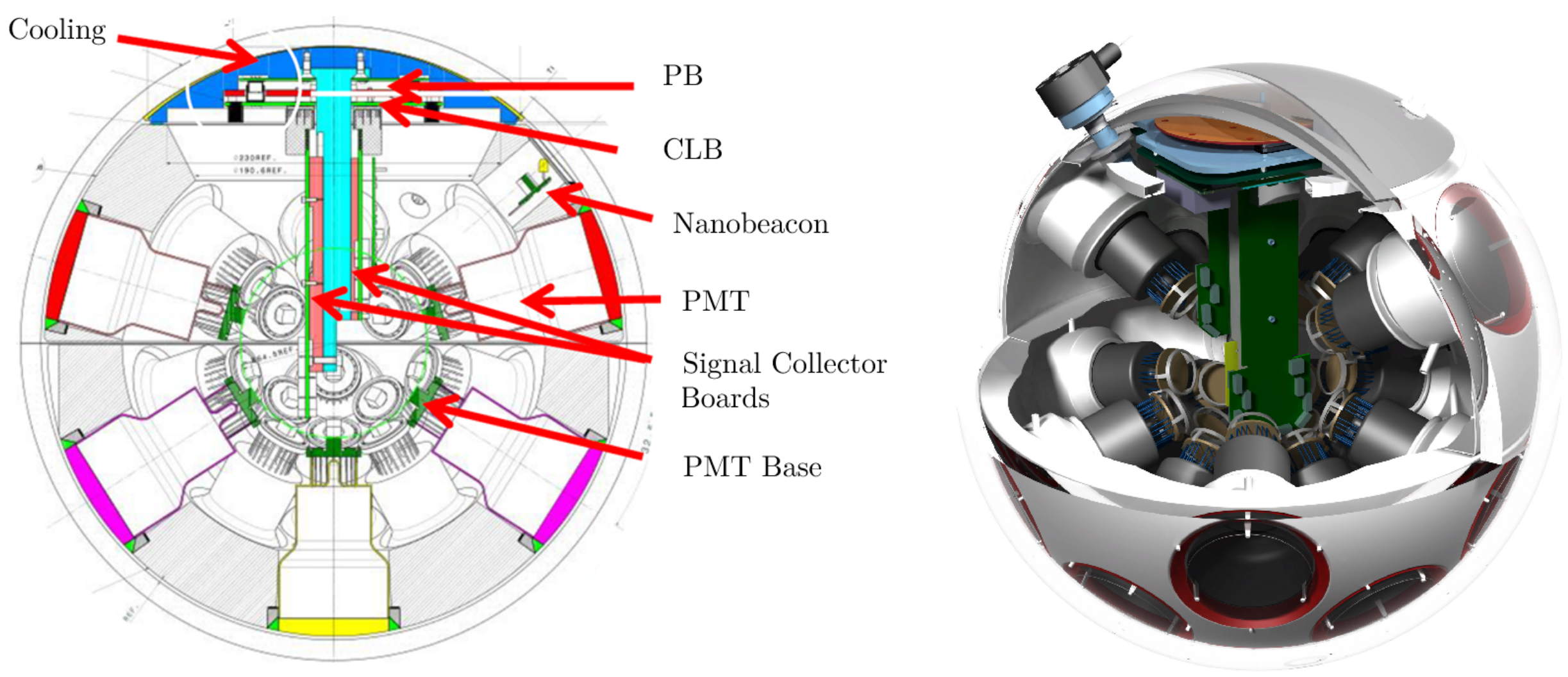

Figure 2: Left: 2D vertical cross-section of the DOM showing the position of the PB and the main elements of the DOM indicated with arrows. Right: 3D representation of the DOM.

## 2. Design and architecture of the Power Board

The PB (Figure 3) is a crucial component of the KM3NeT detector, as it provides power to the CLB and to the rest of the DOM elements. Because of the remote location of KM3NeT in the deep sea, the PB design goals for the power board were low power consumption and high reliability. This is crucial because maintenance of the deployed structures involves complex and costly operations. The architecture of the PB, including its various functionalities, is shown in Figure 4. The PB is supplied with 12 V, which are generated with a DC/DC converter from 400 V at the breakout box, a waterproof enclosure at the input of the DOM that can withstand the harsh conditions of the deep sea. The breakout box is connected, via the DU backbone, to the DU base to receive the high voltage. The main blocks of the PB are:

- The filter block at the input voltage of the PB, which removes the high-frequency noise generated by the power converters at the DU base and breakout board,
- the hysteresis block, which prevents the PB from entering in an unstable state during startup and shutdown,
- the startup block, which allows the different voltages to startup monotonically as needed by the FPGA of the CLB,
- the Nanobeacon power supply controller, which controls the power supply of the Nanobeacon and can be configured through $I^2C$,
- the monitoring system, which reads out the voltages and currents of every power rail, in addition to the temperature sensor.

The main characteristics and functionalities of the PB blocks are detailed in the following subsections.

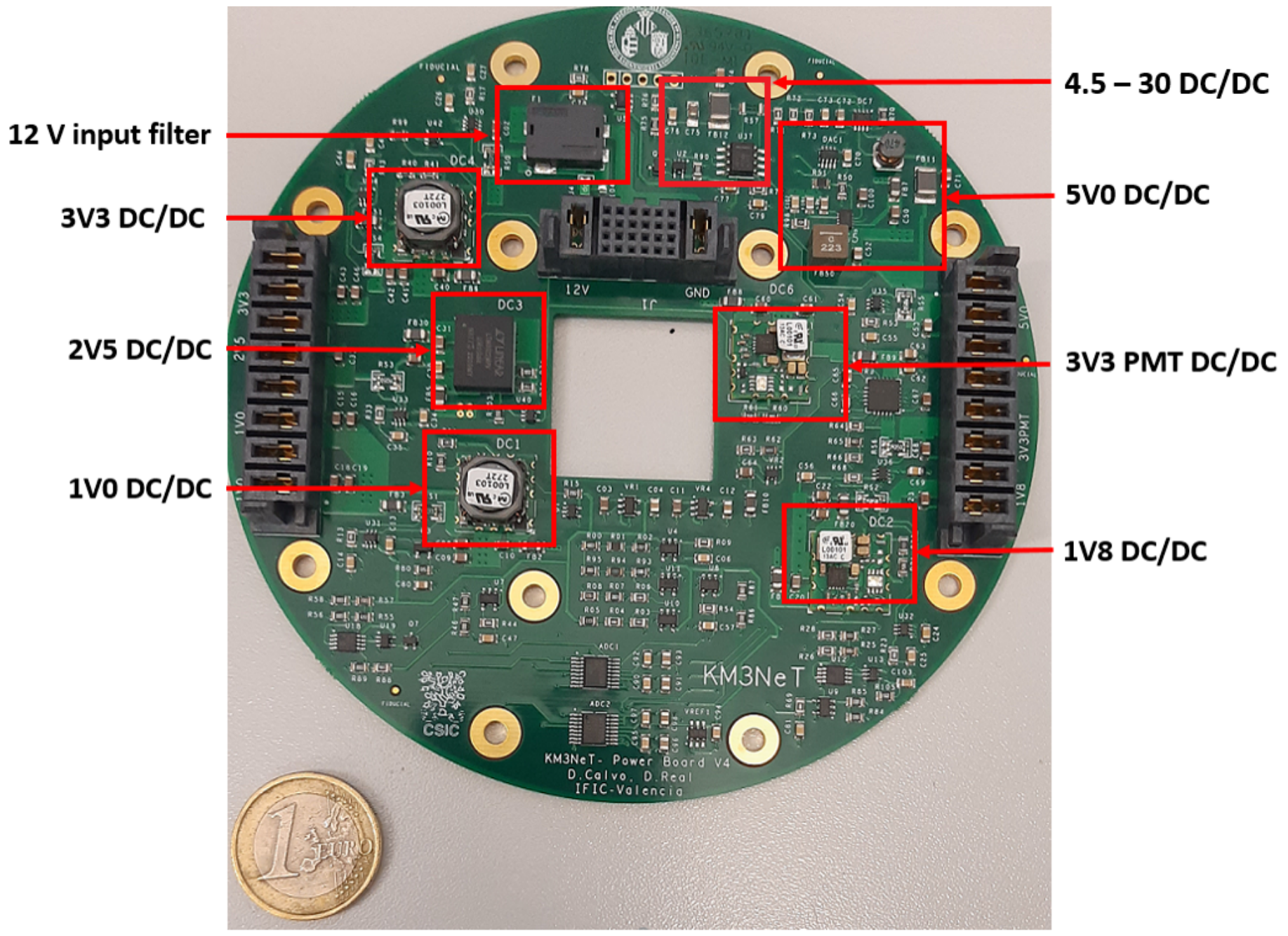

Figure 3: View of the PB of the DOM (upgraded version). The different DC/DC converters to generate the voltages needed by the FPGA and the remaining components of the CLB are marked.

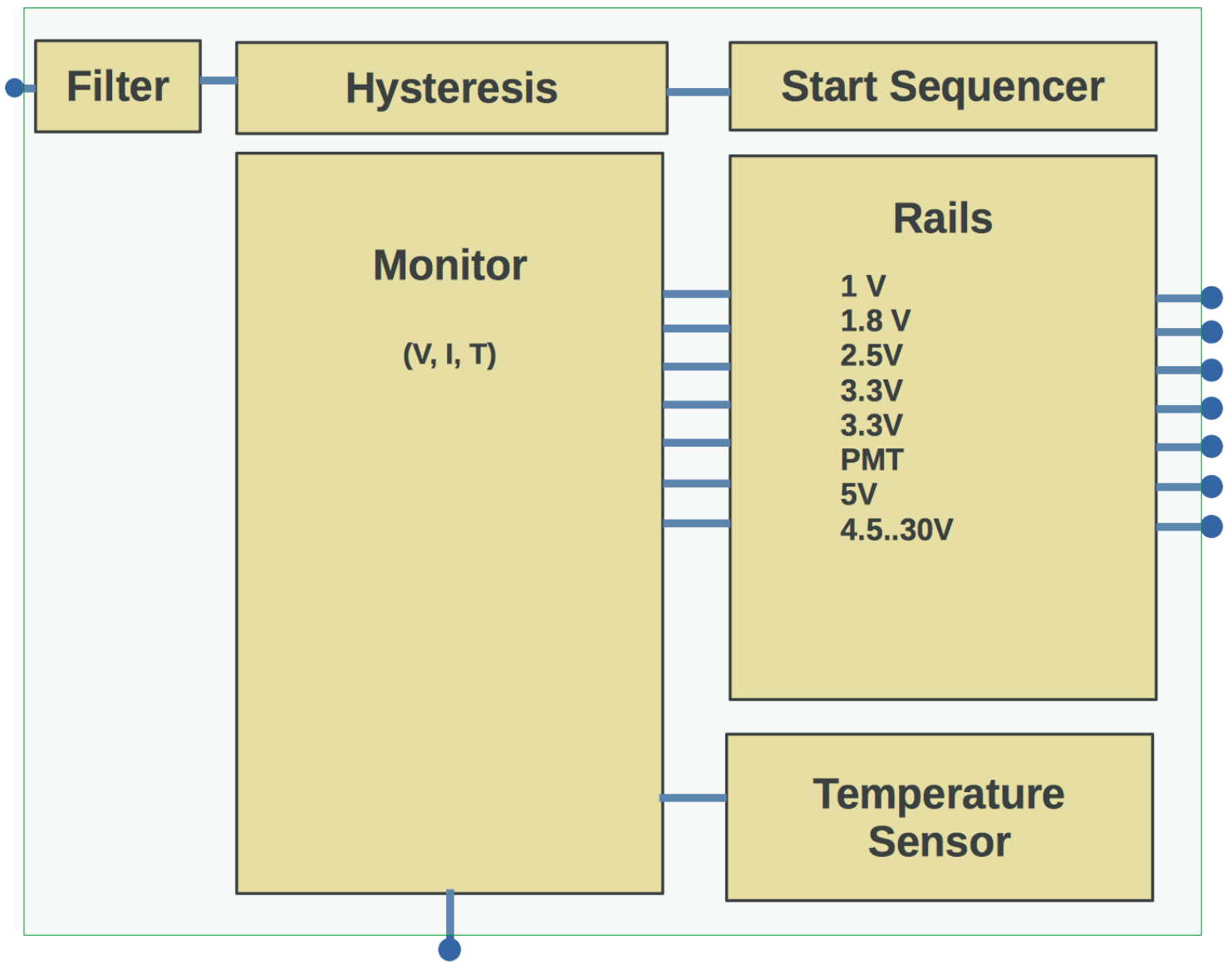

Figure 4: Architecture of the PB. The rails, which provide the different power supplies needed by the DOM, are managed by the start sequencer, which generates at startup the monotonic power sequence requested by the CLB FPGA. The monitor subsystem surveys the voltages and currents of the different rails, as well as the temperature sensor installed on the board. The 12 V are filtered at the input and the hysteresis system prevents instabilities while powering up and down the PB.

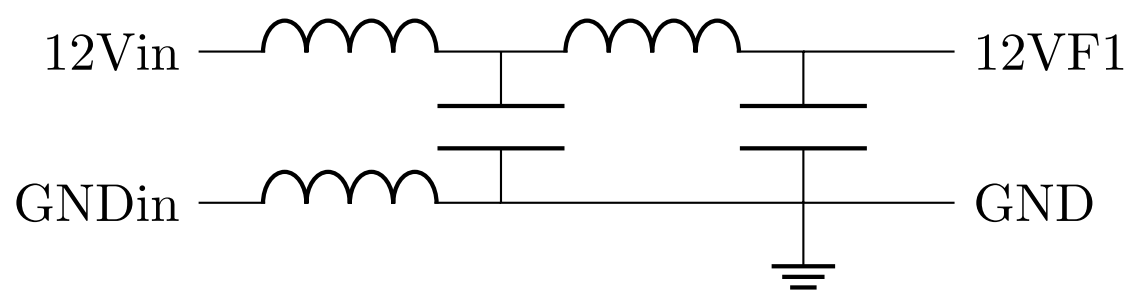

Figure 5: Scheme of the Pi filter functioning as input high-frequency filter on the PB.

### *2.1. Input high-frequency filter*

The PB incorporates an Electro-Magnetic Interference (EMI) filter to remove high-frequency noise from the 12 V DC input signal, providing a low impedance path for the noise, and allowing the desired signal to pass through with minimal attenuation. The filter consists of a Pi filter (see Figure 5) with two capacitors connected to the ground and an inductor connected between the two capacitors. Two inductors connected between the power terminals and the inputs of the Pi filter are also included. The insertion losses of the filter have a minimum of 35 dB from 1 MHz to 1 GHz with a maximum drop of 30 mV. The filter can withstand a maximum voltage of 125 V (DC) and has a rated voltage of 50 V (DC) with a nominal operation of 12 V. The bode diagram of the filter is shown in Figure 6. Overall, the use of the filter improves the performance and reliability of the PB by reducing the impact of EMI noise on its operation.

### *2.2. Hysteresis*

The PB implements a hysteresis loop to avoid instabilities at the startup. The DC/DC converters of the PB are enabled only when the input voltage exceeds 11 V, and are disabled when the input value drops below 9.5 V. In this way, the fluctuations in the PB regulators are avoided during power on/off. The scheme of the hysteresis subsystem is presented in Figure 7. To provide hysteresis functionality, an operational amplifier is used in comparator mode with positive feedback. The input voltage is compared to a reference voltage (3.698 V) and the output is either high or low, depending on whether the input is above or below the reference. The input voltage is supplied to the operational amplifier after a voltage divider circuit, enabling a range of operation from 0 to 5 V. The non-inverting input of the operational amplifier is connected to the operational amplifier output through a resistor. The inverting input of the operational amplifier is connected to the input reference voltage. The resistor values in the voltage divider circuit at the input are chosen to set the reference voltage to the desired switching thresholds. When the input voltage is between 9.5 V and 11 V, the output will be either low (active) or high depending on the current state of the system. If the output is low, the reference voltage will be 9.5 V, and the system will switch off when the input voltage drops below 9.5 V. If the output is high (inactive), the reference voltage will be 11 V, and the system will switch on when the input voltage rises above 11 V. This creates the desired hysteresis effect, where the switching threshold depends on the current state of the system avoiding fluctuations in the switching up and down of the system.

### *2.3. Power startup*

One of the functions of the PB is to provide a proper voltage startup sequence for the FPGA of the CLB [1]. To achieve this, the PB includes a sequencer that generates the required sequence of voltages, which starts monotonically as shown in Figure 8. The output of the hysteresis block initiates the sequence by starting the lower voltage rail (1 V). Then, the *power-good* signal of the lower voltage rail is connected in a cascading manner to prevent the higher voltage rails from starting until the previous rail has started successfully. In this way, a monotonically increasing sequence is produced. Two *power-good* signals are sent outside by the PB for monitoring and control purposes. The first one indicates that the voltage for PMTs has been successfully started (*power-good* PMT). The second one indicates the completion of the entire power-up sequence when the 5 V starts successfully.

[1] `https://docs.xilinx.com/v/u/en-US/ds182_Kintex_7_Data_Sheet`

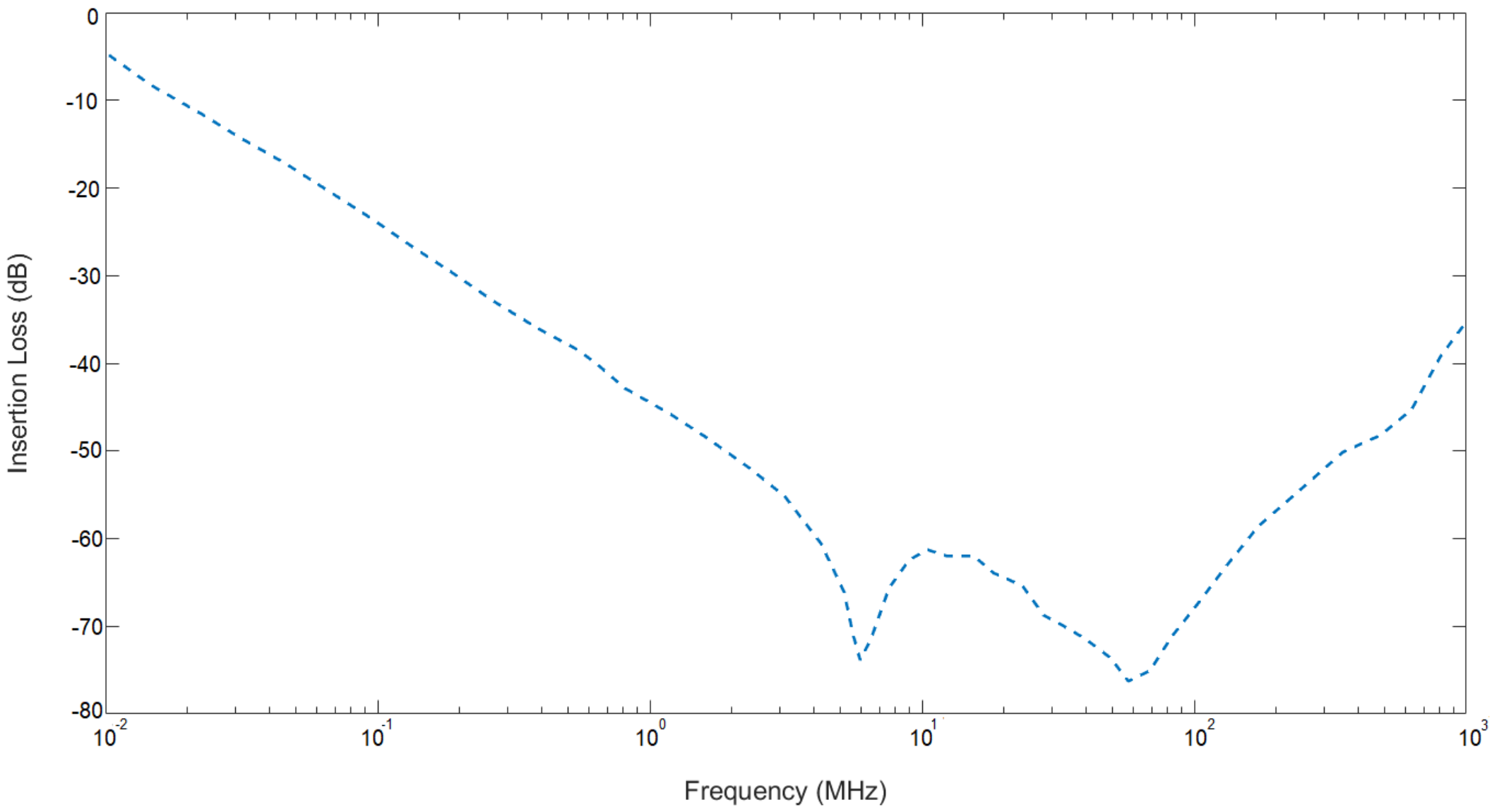

Figure 6: Bode diagram of the Pi filter at the PB to filter out high-frequency noise. From 1 MHz up to 1 GHz the insertion losses are below -35 dB.

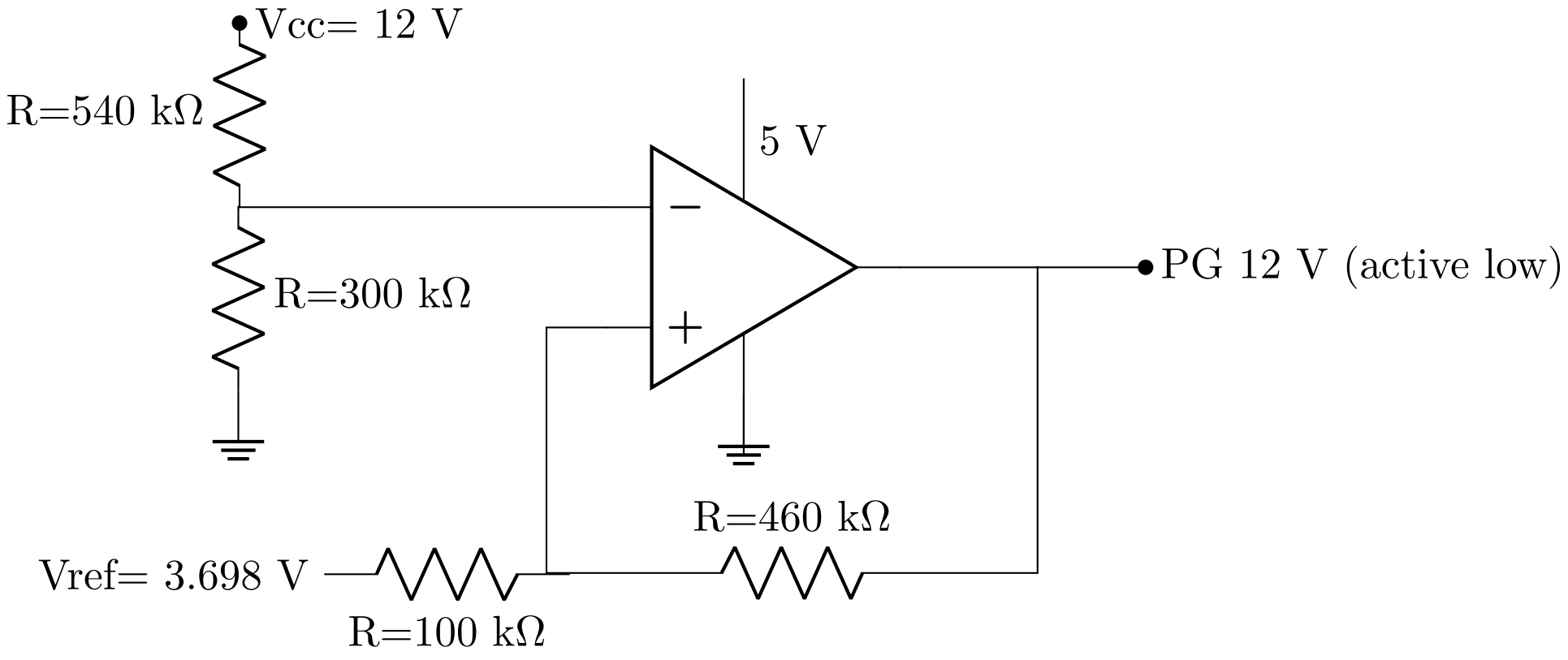

Figure 7: Scheme of the hysteresis subsystem. The configuration of the operational amplifier allows to start at 11 V and to disconnect when the input voltage drops below 9.5 V. In this way, instabilities are prevented at power up and power down.

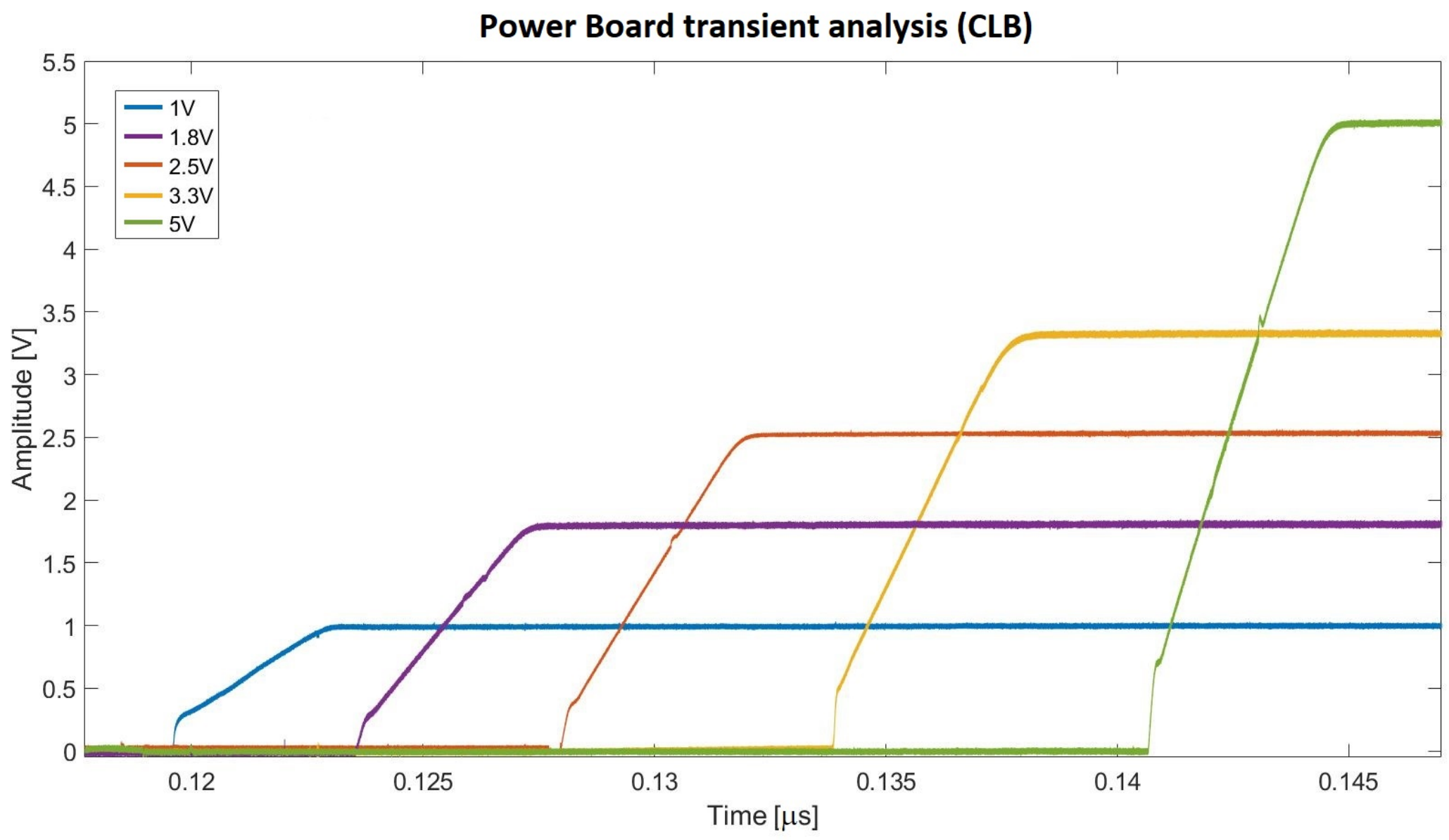

Figure 8: Startup sequence of the PB. The figure shows that the PB indeed generates the various voltages in the sequence needed by the Xilinx FPGA on the CLB.

Table 1: PB efficiency for each rail output for both the original and upgraded versions.

| Voltage (V) | Current (A) | Efficiency original PB (%) | Efficiency upgraded PB (%) |
|---|---|---|---|
| 1 | 0.13 | 80 | 80 |
| 1.8 | 0.33 | 80 | 80 |
| 2.5 | 0.33 | 60 | 78 |
| 3.3 | 0.81 | 65 | 90 |
| 3.3 PMT | 0.46 | 90 | 90 |
| 5 | 0.10 | 60 | 90 |

### *2.4. DC/DC rails*

The PB generates six regulated voltages (1 V, 1.8 V, 2.5 V, 3.3 V, 3.3 V PMT, and 5 V) from the 12 V input using five non-isolated point-of-load (POL) DC/DC converters and a linear regulator. The modular POL approach is easy and fast to implement, and leads to a compact design with a small footprint and a simple Printed Circuit Board (PCB) layout. These designs are optimized by the manufacturer for size, heat flow, and EMI protection. They are also reliable in terms of initial assembly errors, response to fault conditions, and component lifetime failures. In addition, they include sophisticated protection mechanisms like low-pressure molding or chemical protection to avoid corrosion or moisture. The PB employs high-efficiency DC/DC converters to minimize power consumption in the DOM with the exception of the 3.3 V PMT linear regulator used to reduce noise on the PMT voltage rail. The efficiencies of these DC/DC converters are listed in Table 1 for both the original and upgraded versions.

### *2.5. Nanobeacon*

The PB has a configurable output which is used by the Nanobeacon, a time calibration device mounted in the DOM. This configurable output consists of a DC/DC converter that operates in a buck-boost configuration and can provide a voltage that is adjustable between 4.5 V and 30 V through an $I^2C$-controlled 10-bit DAC. When the output voltage of the DAC changes, the control voltage of the DC/DC converter is modified, and the voltage of the Nanobeacon power rail is adjusted accordingly. This voltage determines the amount of current supplied to the LED and, therefore, the intensity of the generated optical pulse.

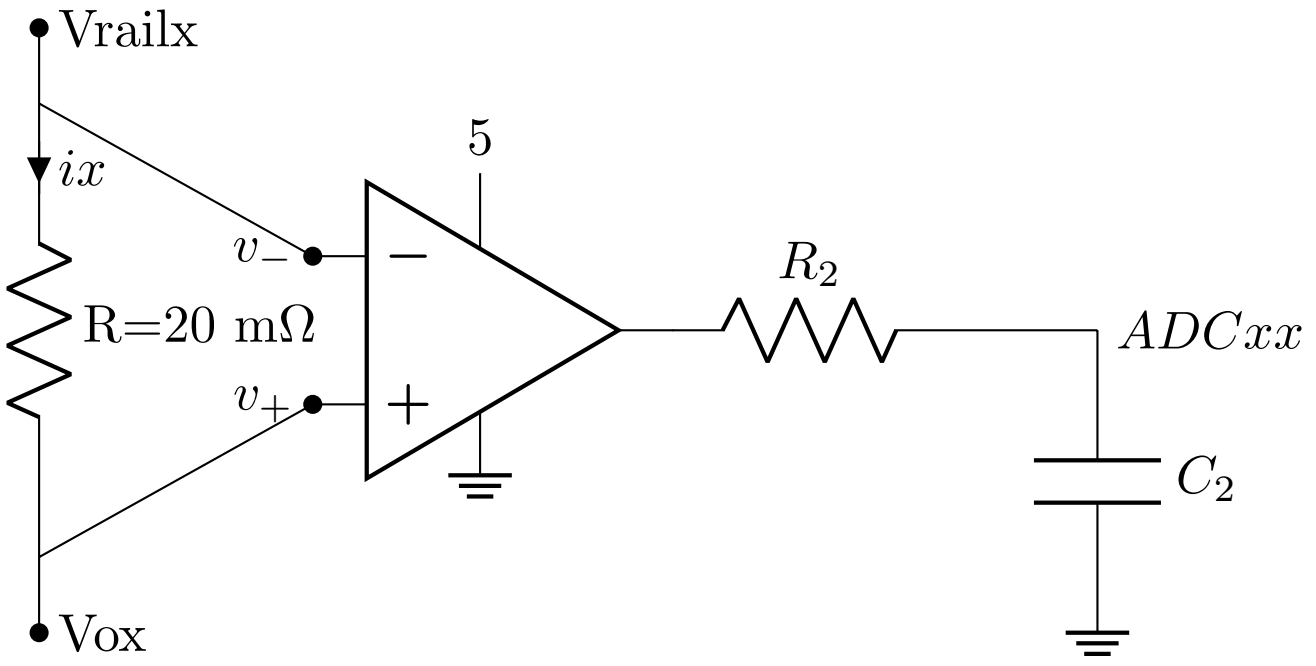

Figure 9: Template of the circuit to read out the current. The output line of a power rail passes through a 20 mΩ resistor, where the drop voltage is amplified in a high-precision amplifier. The output of the amplifier is read out in an ADC channel, and sent via $I^2C$ outside of the PB to the CLB.

### *2.6. Monitoring system*

The currents and voltages of the PB are monitored in real time using two 12-bit ADCs of 12 inputs mounted on the board. In order to read out the current, a resistance of low value, 20 mΩ, is mounted in series in the rail where the current is measured. The two pads of the resistor are connected to a high-precision current-sense amplifier, also called current-shunt monitor, with a fixed gain of 50. As an example, a 0.5 A DC current causes a voltage drop of 10 mV at the 20 mΩ shunt resistor, which is amplified to 0.5 V by the operational amplifier. This voltage is digitized by an ADC channel, which is read out via $I^2C$. Figure 9 shows the design of the circuit used to read out the rail intensity. Some rails, such as the 1.8 V, 2.5 V, and the regulated 5 V, use a shunt resistor of 50 mΩ because the current circulating is lower than in the other rails. The conversion ratio (voltage to current) for each rail is taken into account by the monitoring software. The PB includes also a temperature LM45BIM3 sensor[2], which is read out by one of the ADC channels. The voltages are adapted to the ADC voltage scale and connected to a voltage follower for readout. Voltages, currents, and temperature are read using two MAX1239 ADCs[3]. In Table 2, the different ratios are shown.

### *2.7. Layout*

The layout of the PB has four layers; two for signals, placed on the top and bottom layers, one for power planes, and one for ground. The layer distribution and the stackup chosen for the PB are shown in Figure 10. The bottom of the board has no components so a thermal interface pad can be placed between it and the aluminum frame onto which it is mounted.

### *2.8. Firmware*

The CLB firmware reads the PB's voltages, currents, and temperature. The CLB firmware is a combination of gateware and embedded software, with the gateware made up of logic coded in Hardware Description Language (HDL) and the embedded software written in C [12]. The gateware includes two LatticeMico32 (LM32) microprocessors [4]: the *White Rabbit* processor to manage the optical link traffic and tuneable oscillators, and the other to manage the communication interfaces for the instrumentation devices. The software is organized in three distinct layers - Common, Platform, and Application - where the Application layer holds code for detectors, peripherals, and slow control. This layer manages the ADCs and temperature sensor of the PB.

[2] https://www.ti.com/lit/gpn/lm45

[3] `https://www.analog.com/media/en/technical-documentation/data-sheets/MAX1236-MAX1239M.pdf`

[4] `https://www.latticesemi.com/en/Products/DesignSoftwareAndIP/IntellectualProperty/IPCore/IPCores02/LatticeMico32.aspx`

Table 2: Conversion ratios and associated process variable. The ratios are used to convert the voltage value read by the ADC channel either to Volts or Amperes. In the case of the current readout, the values of the shunt resistance are provided too.

| Process Variable | Ratio | Resistance | Description |
|---|---|---|---|
| 12V current | 1 | 20 mΩ | 12V current |
| 1V0 current | 1 | 20 mΩ | Current at the 1 V rail |
| 1V8 current | 0.4 | 50 mΩ | Current at the 1.8 V rail |
| 2V5 current | 0.4 | 50 mΩ | Current at the 2.5 V rail |
| 3V3 current | 1 | 20 mΩ | Current at the 3.3 V rail |
| 5V0 current | 0.4 | 50 mΩ | Current at the 5 V rail |
| 3V3PMT current | 2 | 10 mΩ | Current at the 3.3 V rail for the PMTs |
| VLED current | 2 | 10 mΩ | Current at the Nanobeacon rail |
| VLED voltage | 10.1 | | Voltage at the Nanobeacon rail |
| 1V0 voltage | 1 | | Voltage at the 1 V rail |
| 1V8 voltage | 1 | | Voltage at the 1.8 V rail |
| 2V5 voltage | 1 | | Voltage at the 2.5 V rail |
| 3V3 voltage | 2 | | Voltage at the 3.3 V rail |
| 5V0 voltage | 2 | | Voltage at the 5.0 V rail |
| 3V3PMT voltage | 2 | | Voltage at the 3.3 V rail for the PMTs |
| VLED control voltage | 1 | | Voltage at the Nanobeacon rail |
| PB Temp | 100 | | Temperature in the PB. Value in °C |

## 3. Power board upgrade

In 2017, an upgrade of the PB was initiated with the aim of improving its overall efficiency. The PB was modified to operate more efficiently by replacing some of the DC/DC converters. These modifications reduce the power consumption, thermal losses, and temperature inside the DOM, thereby increasing the overall reliability of the DOM electronics. A set of new DC/DC converters was chosen paying particular attention to their efficiency at the PB's operating point and to their reliability. The original Murata DC/DC converters [5] of the 2.5 V, 3.3 V, and 5 V rails were changed. For 3.3 V a similar DC/DC model from Murata with the proper current output was selected. For 2.5 V and 5 V, new models from Analog Devices were chosen. After implementing these modifications, a decrease of more than one Watt in the power consumption of the DOM acquisition electronics was achieved (see Table 1).

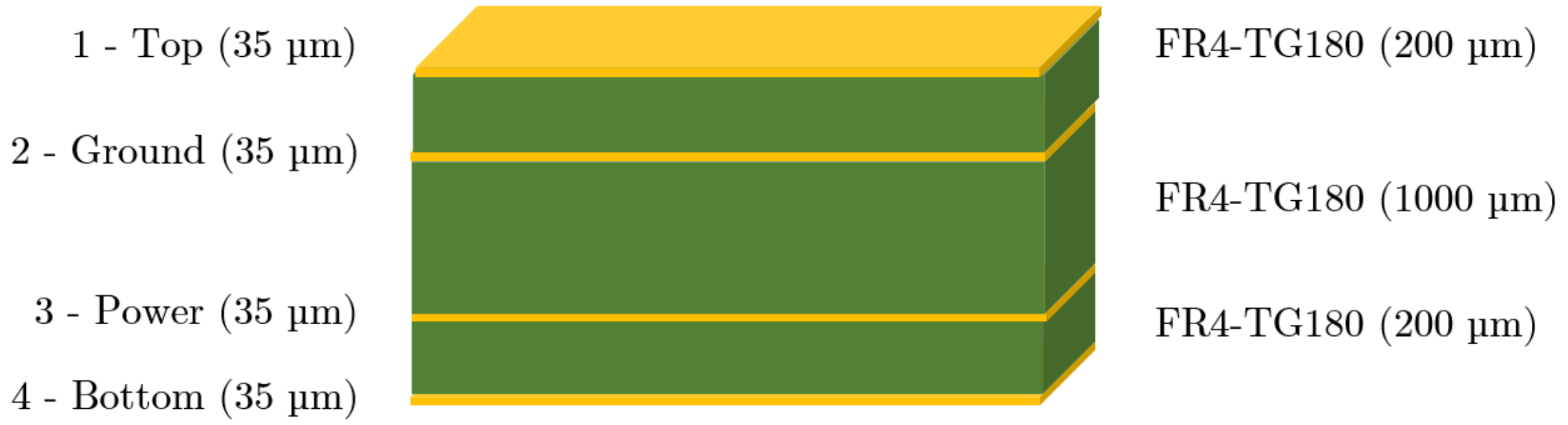

Figure 10: Stackup of the PB PCB. It contains four layers, all of them of copper and with a width of 35 µm. The dielectric material is FR4, with a core of 1,000 µm and two external frames of 200 µm. For a better representation, the image is not to scale.

[5] https://www.murata.com/products/productdata/8807038189598/okl-t3-w12.pdf?1583754815000

## 4. Reliability: FIDES and HALT

The reliability of both versions of the PB has been evaluated using a FIDES analysis [13], and in the case of the upgraded version also by using a Highly Accelerated Life Test (HALT) procedure. These methods are used by KM3NeT to increase the reliability of the boards in the early stages of the development process.

### *4.1. FIDES*

FIDES is the method selected by KM3NeT Collaboration to assess the reliability of electronic boards [14, 15, 16]. It provides a handbook for predicting the reliability of components and a guide for auditing the manufacturing process. Using this method, it is possible to compute an estimate of the Failure In Time (FIT, given in failures per $10^9$ hours) or the Mean Time Between Failure (MTBF) of the analyzed board. The FIDES method takes into account the expected operational conditions or stress, the life profile, and the technological factors that affect the reliability of the board. In addition to the estimated FIT, the method can identify weak points in the design of the board at a very early stage, saving time and costs in the development process for electronic boards. The reliability of the PB has been evaluated providing a FIT value of 947, while the upgraded version has increased its reliability up to a FIT value of 783. This FIT number means that around 90% of the PB will not have any issues during the total KM3NeT operation time. The value calculated refers to the complete board, but there are subsystems that are not critical, such as the piezo sensor or the Nanobeacon power subsystem, the loss of which would not harm the operation of the DOM, so the expected failure rate of boards after the total KM3NeT operation time will be lower than 10%.

### *4.2. HALT*

The HALT method is used to assess the reliability of electronic boards by applying various forms of stress to a small number of boards, usually 4 to 6, at an early design stage. HALT tests are implemented by putting the PBs under extreme temperatures and under extreme rates of temperature change (1 °C/min). The goal of these tests is to ensure the functionality of the product and to optimize the test setup for maximum functional test coverage. The test setup should also allow for remote operation outside the environmental chamber. In KM3NeT, this approach has been introduced and has been used on the upgraded version of the PB. The temperature step stress tests involve decreasing and increasing the temperature of the boards in steps, while the extreme temperature stress tests involve rapid changes in the temperature to the minimum and to the maximum. A total of 6 PBs have undergone HALT tests in combination with 6 CLBs. The minimum temperature reached by the PBs was $-40$ °C, the limit of the climatic chamber used. The maximum temperature was 95 °C, the temperature at which the PB still worked. The tests were stopped when the FPGA had to be switched off as a precaution because the internal temperature was over its operational limit. The results of the HALT tests allow to set the limits for the Highly Accelerated Stress Screening (HASS) tests, which are use during mass production to filter infant mortality.

## 5. Production control

Several thousand PBs have to be produced for the construction of the DOMs and DUs to be deployed at the bottom of the Mediterranean Sea. To ensure their reliability, a series of production requirements have been established:

- The PCB production and assembly process must comply with standard IPC Class 3 [6],

[6] https://www.lab-circuits.com/uploads/doc/tech/Capabilities_for_Class_3_and_3A_IPC_circuits_EN.pdf

- solder paste masks should be generated using the given *gerber files*, choosing the pad shrinking factor based on the solder paste and the mask thickness,
- solder paste must be deposited on the PCB using automatic machines and the aforementioned masks for good uniformity,
- solder paste deposition must be inspected before the PCB is populated,
- all Surface Mounted Device components must be placed using automatic pick&place machines,
- a reflow oven must be used for soldering the components,
- the boards must be identifiable. If the PCBs also have an individual identifying code from the producing company, an electronic file with the correspondence between the board label and the PCB label must be provided,
- the production must provide traceability of all procured components in accordance with IPC1782 [7] level 2 (M2), with level 3 (M3) traceability as a second option.

After production, a series of tests are required as part of the procurement process. The data from these tests is stored in electronic format. Finally, appropriate packaging for shipment to the integration sites should be implemented.

*5.1. PCB Test Control*

Before the PBs are assembled, the following tests are carried out on the PCBs:

- 100% electrical continuity tests
- control of correspondence to IPC Class 3 on a sample of boards performing metalographic micro sections.

*5.2. Component Assembly Test Control*

The following activities are carried out during the assembly of components on the PCB:

- Identification of the board with an appropriate label,
- automatic optical inspection on the positioning and soldering of components on all boards,
- X-Ray inspection and verification of Very-thin Quad Flat Non-leaded package components.

## 6. Functional Tests

Functional tests are performed immediately after the production of the boards to check their correct behavior and identify faulty boards. During these tests, the PB is connected to a CLB. The CLB is programmed to provide a maximum voltage of 30 V on the Nanobeacon device.

The following actions are required for the tests:

- Set the input voltage to 12 V and power-on the system,
- verify that the rail voltages remain within the specified accuracy ranges (see Table 3), as measured at the positions of the CLB indicated in Figure 11,
- write down the measurements after 1 minute and 5 minutes from power-on, and finally, once the test are finished and the boards are powered off, the results of the tests are stored in an electronic file.

[7] https://www.ipc.org/TOC/IPC-1782.pdf

Table 3: Nominal values of voltage for each rail and the percentage of variation allowed at the functional test.

| Rail | Voltage (V) | Accuracy (%) |
|---|---|---|
| 1V0 | 1.0 | ± 1.5 |
| 1V8 | 1.8 | ± 3.0 |
| 2V5 | 2.5 | ± 3.0 |
| 3V3 | 3.3 | ± 1.5 |
| 3V3PMT | 3.3 | ± 3.0 |
| 5V0 | 5.0 | ± 3.0 |
| VLED | 4.0 − 30.0 | ± 1.0 |

## 7. Production and Reliability Results

According to the production control process outlined in the previous sections, a total of 1,750 PBs have been produced in eight different batches. The production yield has been very high (> 99%), with only a few malfunctioning PBs detected during the production tests.

Several hundreds of the produced PBs have been mounted and tested in their corresponding DOMs. During this process, only a few minor issues related to the component assembling process on the PCB were found and this experience was used to improve the packaging and handling instructions.

## 8. Conclusions

The architecture of the KM3NeT power board, together with the different functional blocks has been presented. The power boards will be used for more than a decade in conditions where access and maintenance are very difficult, making their efficiency and reliability crucial. To ensure the quality and reliability of these boards, specific requirements have been established. The present work outlines the measures taken by KM3NeT Collaboration to enhance the reliability of the power boards during production. Additionally, a test bench has been implemented to filter any non-functional boards after production. A total of 6,000 of these boards will be produced for the completion of the KM3NeT infrastructure. As of October 2023, 1,200 power boards were successfully produced with a high yield. A total of 828 power boards are already working in the 46 Detection Units deployed at this time.

## 9. Acknowledgements

The authors acknowledge the financial support of the funding agencies:

Agence Nationale de la Recherche (contract ANR-15-CE31-0020), Centre National de la Recherche Scientifique (CNRS), Commission Européenne (FEDER fund and Marie Curie Program), LabEx UnivEarthS (ANR-10-LABX-0023 and ANR-18-IDEX-0001), Paris Île-de-France Region, France;

Shota Rustaveli National Science Foundation of Georgia (SRNSFG, FR-22-13708), Georgia;

The General Secretariat of Research and Innovation (GSRI), Greece;

Istituto Nazionale di Fisica Nucleare (INFN),Ministero dell'Università e della Ricerca (MIUR), PRIN 2022 program (Grant PANTHEON 2022E2J4RK);

Ministry of Higher Education, Scientific Research and Innovation, Morocco, and the Arab Fund for Economic and Social Development, Kuwait;

Nederlandse organisatie voor Wetenschappelijk Onderzoek (NWO), the Netherlands;

The National Science Centre, Poland (2021/41/N/ST2/01177); The grant "AstroCeNT: Particle Astrophysics Science and Technology Centre", carried out within the International Research Agendas programme of the Foundation for Polish Science financed by the European Union under the European Regional Development Fund;

National Authority for Scientific Research (ANCS), Romania;

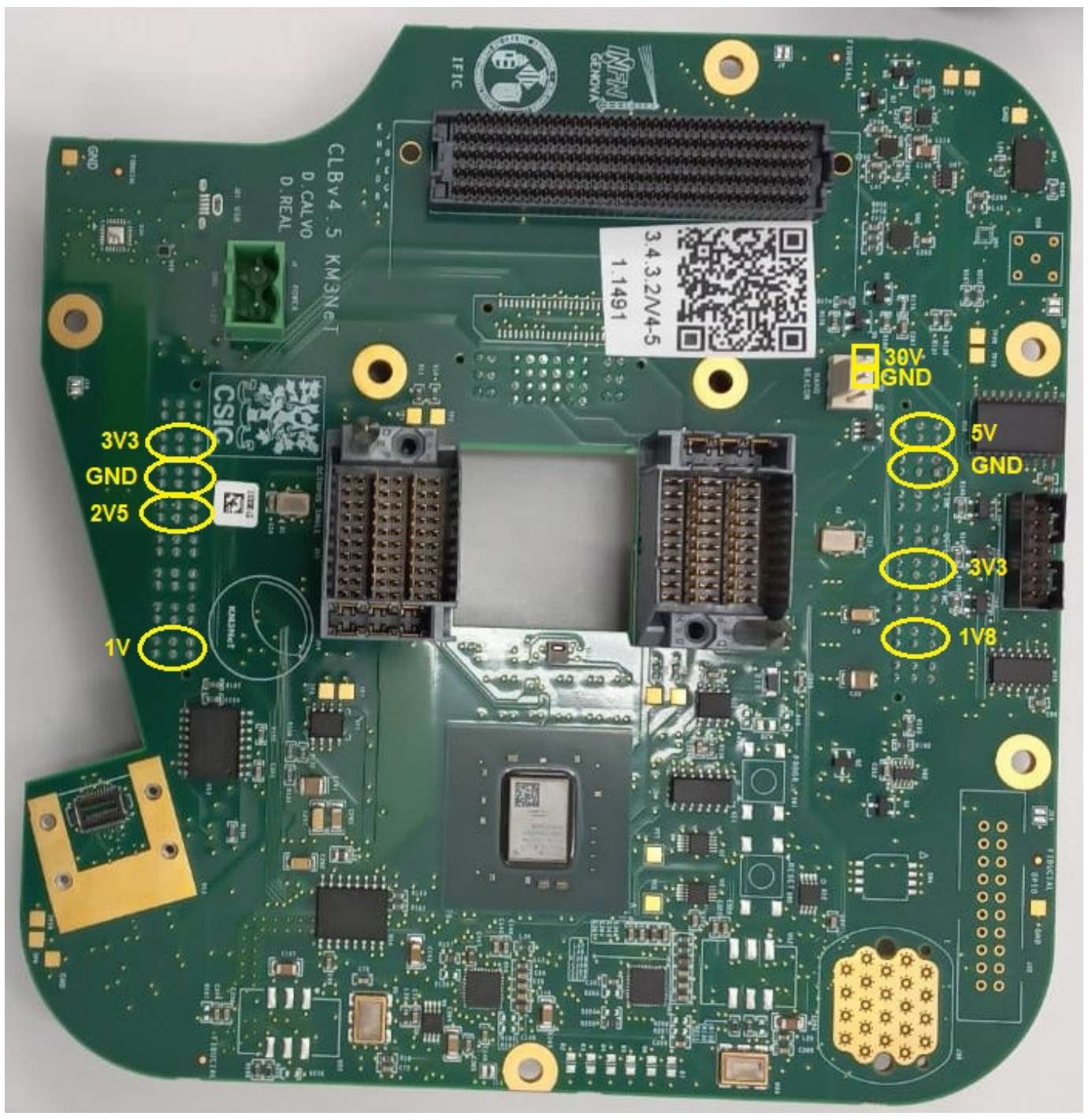

Figure 11: Picture of a CLB with the test points for production functional tests. The different power rail test points are marked on the picture. A CLB running operational firmware is used as load and for measuring the voltages.

Grants PID2021-124591NB-C41, -C42, -C43 funded by MCIN/AEI/ 10.13039/501100011033 and, as appropriate, by "ERDF A way of making Europe", by the "European Union" or by the "European Union NextGenerationEU/PRTR", Programa de Planes Complementarios I+D+I (refs. ASFAE/2022/023, ASFAE/2022/014), Programa Prometeo (PROMETEO/2020/019) and GenT (refs. CIDEGENT/2018/034, /2019/043, /2020/049. /2021/23) of the Generalitat Valenciana, Junta de Andalucía (ref. SOMM17/6104/UGR, P18-FR-5057), EU: MSC program (ref. 101025085), Programa María Zambrano (Spanish Ministry of Universities, funded by the European Union, NextGenerationEU), Spain;

The European Union's Horizon 2020 Research and Innovation Programme (ChETEC-INFRA - Project no. 101008324).

**Diego Real** is a PhD. in Physics and Research Engineer at Instituto de Física Corpuscular. He received his BS in Electronics in 1997 and his MS in Control and Electronics in 2000, both from the Polytechnic University of Valencia. He is the author of several publications on electronics. His PhD. was awarded by Spanish Astronomy Academy with the Prize in the Instrumentation Category. His current research interests include acquisition and synchronisation systems for particle physics. He is, since 2013, the Electronics project leader of the KM3NeT telescope and member of the Technical Advisory Board of the GVD-Baikal telescope.

**David Calvo** is a PhD. in Physics and research engineer at Instituto de Física Corpuscular of Valencia. He received his MS in Computing in 2006 from University Jaume I, his MS in Electronics in 2009 from University of Valencia and his MS in electronic systems design in 2012 from the Polytechnic University of Valencia. His research interests are focused on the digital electronics, synchronization and read out acquisition systems. He is the author of several publications on electronics.